\documentclass{ws-p9-75x6-50}

\begin{document}

\title{Magnetic field tuning of charge and spin order
in the cuprate superconductors}

\author{Anatoli Polkovnikov and Subir Sachdev}

\address{Department of Physics, Yale University,\\
P.O. Box 208120, New Haven CT 06520-8120, USA}

\author{Matthias Vojta}

\address{Theoretische Physik III, Elektronische Korrelationen und
Magnetismus,\\
 Universit\"at Augsburg, 86135 Augsburg, Germany}

\author{Eugene Demler}

\address{Department of Physics, Harvard University, Cambridge MA 02138, USA}


\maketitle

\abstracts{Recent neutron scattering, nuclear magnetic resonance,
and scanning tunneling microscopy experiments have yielded
valuable new information on the interplay between charge and spin
density wave order and superconductivity in the cuprate
superconductors, by using an applied perpendicular magnetic field
to tune the ground state properties. We compare the results of
these experiments with the predictions of a theory which assumed
that the ordinary superconductor was proximate to a quantum
transition to a superconductor with co-existing spin/charge
density wave order.}

\begin{center}
{\tt Invited talk at the conference on `Physical Phenomena at High
Magnetic Fields IV', October 19-25, 2001, Santa Fe, New Mexico}
\end{center}

\section{Introduction}

An innovative series of recent experiments [1-4] have shed new
light on the nature of strong correlations in the cuprate
superconductors. These experiments all use a magnetic field,
applied perpendicular to the CuO$_2$ layers, to tune the low
temperature properties of the superconducting state. Their results
support the idea that ground state correlations in the doped Mott
insulator can be described using a framework of competing order
parameters [5-8]. In our view, they (especially [4]) also offer
compelling evidence that the orders competing with the
superconductivity are spin and charge density waves, and suggest
that one or both are present in the insulating state at very high
fields [9].

In the underdoped regime, experimental evidence for spin and
charge density ordering, coexisting with superconductivity, has
accumulated over the last decade [5,10-13]. With increasing
doping, some experimental results [13,14] can also be explained by
proximity to a quantum critical point at which the spin/charge
order disappears. At optimal doping, it is widely accepted that
the important qualitative characteristics of the ground state are
identical to those of a $d$-wave superconductor described in the
standard BCS-BdG theory; nevertheless even at such dopings an
attractive description of the collective spin and charge
excitations is provided by a theory of the vicinity of the
spin/charge ordering critical point at lower doping [15]. Such a
framework was used recently [16,17] to address the influence of
the applied magnetic field, and a number of its specific
predictions appear to have been confirmed in subsequent
measurements [2-4]. Here we will briefly review aspects of this
theory and discuss extensions needed to develop a complete theory
of the remarkable recent scanning tunneling microscopy (STM)
measurements of Hoffman {\em et al.} [4].

\section{Order parameters and quantum field theory}
We consider a zero temperature transition between a $d$-wave
superconductor (SC) and a superconductor with co-existing
spin/charge order (SC+SDW/CDW). The density wave order parameters
for the transition are defined by the following representations of
spin ($S_{\alpha} ({\bf r},\tau)$, $\alpha=x,y,z$) and charge
density modulation ($\delta \rho ({\bf r},\tau)$) at imaginary
time $\tau$ and on the sites, ${\bf r}$, of a square lattice with
unit lattice spacing:
\begin{equation}
S_\alpha  \left( {\bf r} ,\tau \right) = {\mathop{\rm
Re}\nolimits} \left( {\Phi _\alpha  \left( {\bf r},\tau
\right)e^{i{\bf K}_s  \cdot {\bf r}} } \right){\rm{ ~~;~~ }}\delta
\rho \left( {\bf r},\tau \right) = {\mathop{\rm Re}\nolimits}
\left( {\phi \left( {\bf r},\tau \right)e^{i{\bf K}_c \cdot {\bf
r}} } \right),
\end{equation}
where ${\bf K}_s$ and ${\bf K}_c$ are the respective ordering
wavevectors. A phase with SDW order has $\langle \Phi_{\alpha}
\rangle \neq 0$ and a broken spin rotation symmetry; it is
important to note that we use the term `CDW order' (with $\langle
\phi \rangle \neq 0$) in its most general sense: all local
observables which are invariant under spin-rotation and
time-reversal symmetries (like the on-site energy, or the bond
kinetic and exchange energies, or the bond charge density) acquire
a spontaneous modulation at wavevector ${\bf K}_c$, and it is
possible that the modulation in the total site charge density
itself is quite small. Excluding the case of two sublattice
orderings (with wavevectors $(\pi,\pi), (\pi,0), (0,\pi)$), the
order parameters $\Phi_{\alpha}$, $\phi$ are {\em complex
fields\/}, with their phases representing a sliding degree of
freedom of the density wave. This sliding degree of freedom is
also present when the ordering is commensurate, but the phases
then prefer to take a discrete set of values. We will focus here
on the case ${\bf K}_s=(3\pi/4,\pi)$, ${\bf K}_c=(\pi/2,0)$ which
is experimentally relevant above a doping of about 1/8 (a
theoretical rationale for the selection of this value of ${\bf
K}_c$ was provided in [18]); the generalization to other
commensurate and incommensurate wavevectors is immediate. The
theory also includes order parameters associated with density
waves in the orthogonal direction (${\bf
K}_s^{\prime}=(\pi,3\pi/4)$, ${\bf K}_c^{\prime}=(0,\pi/2)$) but
we will not write down these terms explicitly in the interest of
brevity---this has been done in [17]. A variety of SDW and CDW
phases (with background SC order) are possible in models described
by the fields $\Phi_{\alpha}$ and $\phi$, and some of these have
been described in [19]. Here we will focus on the simplest
possibility of a transition from SC to SC+SDW driven by the
condensation of $\Phi_{\alpha}$ (results are similar in other
cases); as discussed in [19] the condensation of $\Phi_{\alpha}$
also leads to concomitant CDW order with ${\bf K}_c=2{\bf K}_s$
(modulo a reciprocal lattice vector), and near the transition we
can pin the charge order field to the spin order field by
\begin{equation}
\phi \left( {{\bf r},\tau } \right) \propto \Phi _\alpha ^2 \left(
{{\bf r},\tau } \right)
\end{equation}
where a summation over the repeated index $\alpha$ is implied here
and henceforth.

The effective action for $\Phi_{\alpha}$ describing the quantum
transition from SC to SC+SDW is [6,7,15-19]:
\begin{eqnarray}
S_\Phi   = \int d^2 {\bf r} d\tau  && \biggl[ {\left| {\partial
_\tau \Phi _\alpha  } \right|^2 + c_x^2 \left| {\partial _x \Phi
_\alpha } \right|^2 + c_y^2 \left| {\partial _y \Phi _\alpha  }
\right|^2 + s\left| {\Phi _\alpha  } \right|^2  + \frac{{u_1
}}{2}\left( {\left| {\Phi _\alpha  } \right|^2 }
\right)^2 }  \nonumber \\
&&~~~~~~~~~~~~~~~~~ { + \frac{{u_2 }}{2}\left| {\Phi _\alpha ^2 }
\right|^2 + \lambda \left( {\left( {\Phi _\alpha ^2 } \right)^4 +
{\rm{c}}{\rm{.c}}{\rm{.}}} \right)} \biggr]
\end{eqnarray}
where $c_{x,y}$ are velocities, and $s$ is a coupling which tunes
the system across the SC to SC+SDW transition. We have neglected
couplings to the fermionic nodal quasiparticles as we assume they
have been suppressed by constraints from momentum conservation
[16,17]. The non-linearity $u_2$ selects between spiral and
collinear spin orderings [19,17], and we assume $u_2<0$, for which
case collinear ordering with $\epsilon_{\alpha\beta\gamma} \langle
\Phi_{\beta} \rangle \langle \Phi_{\gamma}^{\ast} \rangle=0$ is
selected, as is the case experimentally. The $\lambda$ term is
special to the value ${\bf K}_s=(\pi/4,0)$ and prefers that the
phase of $\Phi_{\alpha}$ take a set of discrete values: it is
permitted for this value of ${\bf K}_s$ because translations by
integer lattice spacings correspond to the discrete transformation
(see (1))
\begin{equation}
\Phi _\alpha   \to \Phi _\alpha  e^{in\pi /4} {\rm{ ,~ }}n{\rm{~
integer,}}
\end{equation}
under which all the terms in $S_{\Phi}$ are invariant; the sign of
$\lambda$ chooses between bond and site centered density waves
[17].

The SC phase with no SDW order is realized for $s$ larger than
some critical value $s_c$. An important property of this phase is
that the $\Phi_{\alpha}$ quanta constitute a stable excitonic
excitation with a {\em 6-fold} degeneracy. At the Gaussian level
this is evident from the fact that the quadratic terms in (3) have
a global O(6) symmetry. However, the terms proportional to $u_2$
and $\lambda$ are only invariant under O(3) spin rotation symmetry
and the discrete symmetry (4), but a perturbative computation to
all orders easily shows that this symmetry is sufficient to
preserve the 6-fold degeneracy. The usual degeneracy of a spin
$S=1$ triplet has been doubled by the additional discrete sliding
degree of freedom of the SDW.

Now let us consider the influence of an applied magnetic field,
$H$, on the SC phase. An important early contribution was made by
Arovas {\em et al.} [20] who focused on cores of the vortices in
the SC order introduced by $H$ and argued that, because of a
microscopic repulsion between the SC and N\'{e}el orders, locally
the SC order would ``rotate'' into an insulating N\'{e}el (SDW)
phase. The cores have since been examined in a number of
experiments [21,22,3] and no clear sign of such behavior emerged:
instead the sub-gap conductance is enhanced in the cores,
additional core states appear, and there may be a subdominant
pairing amplitude ({\em e.g.} $d_{x^2-y^2}+id_{xy}$) [23]. The
nuclear magnetic resonance (NMR) experiments observed enhanced
antiferromagnetic {\em fluctuations} located {\em outside} the
vortex core, as was originally predicted in [16] (see below); the
formalism of [20] allows static antiferromagnetic order on scales
larger than the core size, and this along with the extension to
dynamic antiferromagnetism was discussed recently in [24].

Two of us and Y. Zhang [16] examined the consequences of the
microscopic repulsion between SC and SDW orders, but near a bulk
transition between the SC and SC+SDW phases. We argued that
effects driven by the strongly relevant $u_{1,2}$ interactions in
(3) implied that the predominant enhancement of the SDW
fluctuations in the SC phase, and the consequent lowering of the
exciton energy, occurred primarily in the superconducting region
outside the vortex cores, driven by the superflow induced by $H$.
(Loosely speaking, this can be understood as follows: as we
approach the onset of SDW order by decreasing $s$, $\Phi_{\alpha}$
initially attempts to condense in the cores of the vortices.
However, the quartic self interactions $u_{1,2}$ are most
effective in this strongly localized region, and this drives up
the effective Hartree potential felt by $\Phi_{\alpha}$ in the
core. So $\Phi_{\alpha}$ can only condense in an extended state
which is primarily sensitive to the large spatial region outside
the core over which the superflow is present. Alternatively
stated, the energy of any $\Phi_{\alpha}$ states which may be
localized in the core {\em always} remains at non-zero energies of
order of the exchange interaction (and so are probably strongly
overdamped), and Bose condensation of $\Phi_{\alpha}$ can only
occur in extended states outside the core whose energy can indeed
approach zero [25].) These predictions appear to have been
confirmed in subsequent experiments: the NMR experiments of [3]
were noted above, and the recent STM measurements of [4] observe
CDW modulations (which are related to the SDW by (2)) at distances
almost an order of magnitude beyond the point where the
superconducting coherence peaks are fully recovered outside the
vortex core. The work of [16] also made predictions on the
$H$-dependence of the elastic Bragg peaks associated in the SC+SDW
phase: these are in good accord with subsequent neutron scattering
measurements [2].

We briefly mention the field theory origin of the results of [16].
As we are focusing on long length scales outside the small vortex
cores, a continuum description is possible. For the collective
spin/charge excitations we use $S_{\Phi}$ in (3); for simplicity,
we do not include a Zeeman coupling to $H$ in $S_{\Phi}$ because
it only modifies physical properties at order $H^2$ (because,
loosely speaking, the average field on the spins in the
oscillating SDW vanishes). The infinite diamagnetic susceptibility
of the SC order makes its response to $H$ much stronger, and we
describe this, and the coupling to $\Phi_{\alpha}$, by the action
$S_{\Phi}+S_{\Psi}$ (this cannot be applied within the core where
the SC wavefunction is perturbed in different ways, as noted
above), where
\begin{equation}
S_\Psi   = \int {d^2 rd\tau \left[ {\left| {\left(  {\bf \nabla} -
i {\bf A} \right)\Psi } \right|^2  - \left| \Psi  \right|^2  +
\frac{{\left| \Psi  \right|^4 }}{2} + \kappa \left| \Psi \right|^2
\left| {\Phi _\alpha  } \right|^2 } \right]},
\end{equation}
$\Psi({\bf r})$ is the superconducting order parameter which is
dependent on ${\bf r}$ but independent of $\tau$, we have chosen
various scales to make many couplings in (5) unity [16], ${\bf A}$
is the vector potential of the applied field, and $\kappa>0$ is
the repulsive coupling between the two orders. The action
$S_{\Phi}+S_{\Psi}$ provides a theory of the SC to SC+SDW
transition, including the ingredients for the repulsive
interactions between the excitons and for the lowering of the
exciton energy by the superflow.

In the context of the application to the recent STM measurements
[4], a significant property of $S_{\Phi}+S_{\Psi}$ is that as long
we are in the SC phase (which is evidently the case in [4]), not
only do we have no static SDW order with $\langle \Phi_{\alpha}
\rangle =0$, but we also have $\langle \phi \rangle = \langle
\Phi_{\alpha}^2 \rangle =0$, and so there is no local static CDW
order even in an external field, even though the energy of the
spin/charge exciton has been considerably lowered. This is simply
a consequence of the fact that all the terms in (3) and (5) are
invariant under the sliding symmetry (4), and so the phase of the
exciton has not been pinned. However, the presence of the vortices
clearly breaks the translational symmetry on the lattice scale,
and so the continuum theory should be supplemented by additional
terms which implement this effect and pin the exciton; a little
thought using (1) shows that the simplest such term is
\begin{equation}
S_{{\rm{lat}}}  =  - \zeta \int {d\tau \sum\limits_{\bf r} {\left|
{\Psi \left( {\bf r} \right)} \right|^2 \Phi _\alpha ^2 \left(
{{\bf r},\tau } \right)e^{i 2{\bf K}_s  \cdot {\bf r}}  +
{\rm{c}}{\rm{.c}}.} }
\end{equation}
where $\zeta$ is a new coupling constant. Because of the rapidly
oscillating term, (6) will vanish in any region of space where
$\Psi({\bf r})$ is slowly varying. So the pinning of the exciton
by (6) happens mainly in the vortex core (the diameter of the core
is of the order of a couple lattice spacings), while the energy of
the exciton is lowered by $\kappa$ term in (5) by the superflow
outside the vortex core. On long scales outside the vortex core it
should be acceptable to replace (6) by
\begin{equation}
S_{{\rm{lat,1}}}  =  - \zeta_1 \int {d\tau \,\Phi _\alpha ^2
\left( {{\bf r} = 0,\tau } \right)e^{i\delta }  +
{\rm{c}}{\rm{.c}}{\rm{.}}}
\end{equation}
for a vortex at ${\bf r}=0$, where the phase $\delta$ is
determined by the microstructure of the vortex on the lattice
scale. The term (7) breaks the symmetry (4), and so it is now
possible to have static CDW order in the SC phase in an applied
magnetic field. We still have $\langle \Phi_{\alpha} \rangle =0$,
but (7) does locally lift the 6-fold degeneracy of the exciton to
3+3. We can compute the static CDW order in the spirit of the
calculation of [16], and in the Gaussian approximation to
$S_{\Phi} + S_{\Psi}+ S_{{\rm lat,1}}$, and to first order in
$\zeta_1$ we find for large $|{\bf r}|$
\begin{equation}
\left\langle {\Phi _\alpha ^2 \left( {{\bf r},\tau } \right)}
\right\rangle  = \left( \frac{3}{8 \pi^{3/2} s_1^{1/4} c^{5/2}}
\right) \zeta_1  e^{ - i\delta }  \frac{e^{-2 r_1 \sqrt{s_1}/c
}}{r_1^{3/2}},
\end{equation}
where $s_1$ is the exciton ``mass'' $s$ renormalized downwards by
the superflow via the $\kappa$ term in (5) (as in [16]), $c=(c_x
c_y)^{1/2}$ and $r_1 = c \left( (x/c_x)^2+(y/c_y)^2
\right)^{1/2}$; this implies a static CDW by (1) and (2). The
length scale, $\xi_c =c/(2 \sqrt{s_1})$, over which this CDW order
appears has been significantly increased by the superflow around
the vortex core {\em and\/} the exciton interactions $u_{1,2}$,
while the order has been pinned by the vortex core via (7). In the
absence of the pinning in (7) we would have $\langle \delta \rho
({\bf r}, \tau) \rangle = 0 $ in this SC phase.

\section{Quasiparticle density of states}

We now discuss the application of the above theoretical framework
to the STM measurements in BSCCO of Hoffman {\em et al.} [4]. As
we indicated earlier, they report observations of a modulation of
period 4 in the local density of states (LDOS) in a small energy
window around $\pm 7$ meV in the superconducting region (with
fully formed coherence peaks) outside the vortex core. Such a
modulation can arise by scattering of the quasiparticles off
either the SDW or the CDW correlations. The SDW order is dynamic
and one of its effects will be an enhancement of the quasiparticle
LDOS at energies large enough to allow the quasiparticles to decay
by emission of a finite energy spin exciton. It remains to be seen
whether the spin excitation spectrum of BSCCO in a magnetic field,
as observed in neutron scattering experiments, has spectral weight
at low enough energies for this effect to be important. We also
note that the modulation in the LDOS will only arise as a
consequence of the 3+3 splitting of the spin exciton by the
potential in (6) or (7). Here we only present our results for the
simpler case of modulation induced directly by the CDW order; this
order is made static by the pinning induced by the vortex cores
(as in (8)), and can serve as an elastic scattering potential for
the quasiparticles. The effect of an isolated elastic scattering
potential in a $d$-wave superconductor has been well studied [26],
and for large scattering a virtual bound state is formed at low
energies; the effect of (8) can loosely be interpreted as a
periodic version of this, with the weaker potential only leading
to a periodic `hump' or `shoulder' in the quasiparticle LDOS.

This proposal immediately raises a puzzle which we have not fully
resolved. In general, symmetry arguments indicate that the
presence of the static CDW order in (8) implies that the
quasiparticle LDOS should display a period 4 modulation at {\em
all} energies, and not just on the sub-gap feature at $\pm 7$ meV.
A possible resolution of this puzzle is provided by an appeal to
the effects of disorder: the signal-to-noise for any periodic
modulation is largest at sub-gap energies, and that is where the
CDW modulation is visible. At higher energies, random fluctuations
in the background LDOS of the $d$-wave superconductor, and
especially in the energy at which the coherence peaks are present,
can mask the presence of a periodic modulation. A further
consequence of our proposal is that the CDW order should be
visible near any strong short distance imperfection which can
provide a pinning potential as in (7), and not just near vortex
cores.

As promised above, we conclude with a simple calculation of the
effect of (8) on the LDOS. We diagonalized a $d$-wave BCS
Hamiltonian on a square lattice,
\begin{equation}
H = \sum_{ij} \left(-t_{ij} c_{i\sigma}^\dagger c_{j\sigma}
             + \Delta_{ij} c_{i\sigma}^\dagger c_{j,-\sigma}^\dagger +
h.c.\right)
 + \sum_i  [v({\bf r}_i) - \mu] c_{i\sigma}^\dagger c_{i\sigma} \,,
\end{equation}
where $t_{ij}$ includes nearest-neighbor ($t$) and next-neighbor
($t'$) hopping, $\mu$ is the chemical potential, the pairing
amplitude $\Delta_{ij}$ is self-consistently determined from
$\Delta_{ij} = V_{ij} \langle c_{i\sigma} c_{j,-\sigma} \rangle$,
and $V_{ij}$ is the nearest-neighbor pairing potential. The CDW
modulation (1), (2), (8) is implemented via $v({\bf r}) = v_1
\{\cos [{\bf K}_c \cdot ({\bf r-r}_0)] + \cos [{\bf K}_c^{\prime}
\cdot ({\bf r-r}_0)]\}\,
              e^{-|{\bf r-r}_0|/\xi_c} \, (|{\bf r-r}_0|^2+1)^{-3/4}$.
In this initial calculation we did not include the usual
modulation of the $\Delta_{ij}$ induced directly by the presence
of the vortex: we expect this to be important only in the core of
the vortex, whose physics we are not attempting to model here, and
where the present BCS model is probably inadequate anyway. Fig.~1
shows a numerical result for the LDOS, integrated over an interval
of sub-gap energies, obtained for a periodic $32 \times 32$ system
with 20\% (bulk) hole doping.
\begin{figure}
\epsfxsize=5in \centerline{\epsffile{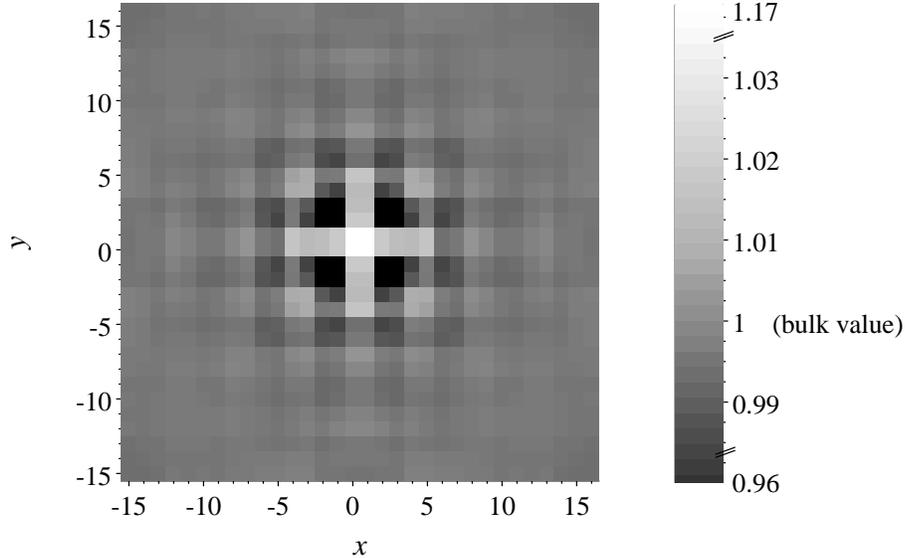}} \caption{
Grayscale plot of the CDW-induced quasiparticle LDOS, integrated
over an energy interval $\Delta_0/10 \ldots \Delta_0/3$, where
$\Delta_0$ is the size of the bulk $d$-wave gap. The CDW
modulation is plaquette-centered at (0.5,0.5), with wavevectors
${\bf K}_c=(\pi/2,0)$ and ${\bf K}_c^{\prime} = (0, \pi/2)$, decay
length $\xi_c=6$, and a strength $v_1 = t/5$. For computational
reasons we have used a large pairing potential, $V=2t$, but we do
not expect qualitative changes for smaller $V$. A similar
modulation is observed at the corresponding energy interval at
negative energies.}
\end{figure}
The charge-density modulation caused by $v({\bf r})$ clearly leads
to a period-4 modulation of the sub-gap LDOS.

\section*{Acknowledgements}
We are especially grateful to Seamus Davis for communicating the
results of [4] prior to publication, and for many useful
discussions. We thank Shoucheng Zhang for valuable comments on the
manuscript, and Steve Girvin for helpful discussions. This
research was supported by US NSF Grant DMR 0098226 and by the DFG
(SFB 484).


\begin{thebibliography}{99}

\bibitem{r1} B. Lake, G. Aeppli, K. N. Clausen, D. F. McMorrow, K. Lefmann,
N. E. Hussey, N. Mangkorntong, M. Nohara, H. Takagi, T. E. Mason,
and A. Schr\"oder, {\em Science} {\bf 291}, 1759 (2001).

\bibitem{r2} B. Khaykovich, Y. S. Lee, S. Wakimoto, K. J. Thomas, R. Erwin,
S.-H. Lee, M. A. Kastner, and R. J. Birgeneau, submitted to {\em
Nature}.

\bibitem{r3} V. F. Mitrovi\'{c}, E. E. Sigmund, M. Eschrig, H. N. Bachman, W. P.
Halperin, A. P. Reyes, P. Kuhns, and W. G. Moulton, {\em Nature}
{\bf 413}, 501 (2001).

\bibitem{r4} J. E. Hoffman, E. W. Hudson, K. M. Lang, V. Madhavan, S. H.
Pan, H. Eisaki, S. Uchida, and J. C. Davis, submitted to {\em
Science}.

\bibitem{r5} J.~Zaanen, {\em Physica} C {\bf 317}, 217 (1999) and references
therein; V.~J.~Emery, S.~A.~Kivelson, and J.~M.~Tranquada, {\em
Proc. Natl. Acad. Sci. USA} {\bf 96}, 8814 (1999) and references
therein.

\bibitem{r6} S. Sachdev and J. Ye, {\em Phys. Rev. Lett.} {\bf 69}, 2411 (1992);
A. V. Chubukov, S.~Sachdev and J. Ye, {\em Phys. Rev.\/} B {\bf
49}, 11919 (1994).

\bibitem{r7} S.-C. Zhang, {\em Science} {\bf 275}, 1089 (1997).

\bibitem{r8} S. Sachdev, {\em Science}, {\bf 288}, 475 (2000) and references
therein.

\bibitem{r9} G. S. Boebinger, Y. Ando, A. Passner, T. Kimura, M. Okuya,
J. Shimoyama, K. Kishio, K. Tamasaku, N. Ichikawa, and S. Uchida,
{\em Phys. Rev. Lett.} {\bf 77}, 5417 (1996).

\bibitem{r10} C. Panagopoulos, B. D. Rainford, J. R. Cooper, C. A. Scott, and
T. Xiang, cond-mat/0007158; C. Panagopoulos, B. D. Rainford, J. L.
Tallon, T. Xiang, J. R. Cooper, and C. A. Scott, preprint.

\bibitem{r11} Y. S. Lee, R. J. Birgeneau, M. A. Kastner, Y. Endoh, S.
Wakimoto, K. Yamada, R. W. Erwin, S.-H. Lee, and G. Shirane, {\em
Phys. Rev.\/} B {\bf 60}, 3643 (1999).

\bibitem{r12} J. E. Sonier, J. H. Brewer, R. F. Kiefl, R. H. Heffner, K.
Poon, S. L. Stubbs, G. D. Morris, R. I. Miller, W. N. Hardy, R.
Liang, D. A. Bonn, J. S. Gardner, and N. J. Curro,
cond-mat/0108479.

\bibitem{r13} P.M. Singer, A.W. Hunt, and T. Imai, cond-mat/0108291.

\bibitem{r14} G. Aeppli, T. E. Mason, S. M. Hayden, H. A. Mook, and J. Kulda,
{\em Science} {\bf 278}, 1432 (1998).

\bibitem{r15} S. Sachdev, C. Buragohain, and M. Vojta, {\em Science}, {\bf 286},
2479 (1999); M. Vojta, C. Buragohain, and S. Sachdev, {\em Phys.
Rev.\/} B {\bf 61}, 15152 (2000).

\bibitem{r16} E. Demler, S. Sachdev, and Y. Zhang, {\em Phys. Rev. Lett.} {\bf
87}, 067202 (2001).

\bibitem{r17} S. Sachdev, cond-mat/0108238.

\bibitem{r18} M. Vojta and S. Sachdev, {\rm Phys. Rev. Lett.} {\bf 83}, 3916 (1999);
M. Vojta, Y.~Zhang, and S. Sachdev, {\em Phys. Rev.\/} B {\bf 62},
6721 (2000).

\bibitem{r19} O. Zachar, S. A. Kivelson, and V. J. Emery, {\em Phys. Rev.\/} B
{\bf 57}, 1422
(1998).

\bibitem{r20} D. P. Arovas, A. J. Berlinsky, C. Kallin, and S.-C. Zhang, {\em
Phys. Rev. Lett.} {\bf 79}, 2871 (1997).

\bibitem{r21} Ch. Renner, B. Revaz, K. Kadowaki, I. Maggio-Aprile, and
\O. Fischer, {\em Phys. Rev. Lett.} {\bf 80}, 3606 (1998).

\bibitem{r22} S. H. Pan, E. W. Hudson, A. K. Gupta, K.-W. Ng, H. Eisaki, S.
Uchida, and J. C. Davis, {\em Phys. Rev. Lett.} {\bf 85}, 1536
(2000).

\bibitem{r23} M. Franz and Z. Tesanovic, {\em Phys. Rev. Lett.} {\bf 80}, 4763
(1998).

\bibitem{r24} J.-P.~Hu and S.-C.~Zhang, cond-mat/0108273.

\bibitem{r25} This perspective differs from that of [24],
whose formulation did not account for the strong effects of the
quartic exciton self-interactions $u_{1,2}$.

\bibitem{r26} A. V. Balatsky, M. I. Salkola, and A. Rosengren, {\em Phys.
Rev.\/} B {\bf 51}, 15547 (1995).

\end{thebibliography}
\end{document}